\documentclass[%
 reprint,
superscriptaddress,
 amsmath,amssymb,
 aps,
prl,
]{revtex4-2}

\usepackage{graphicx}
\usepackage{dcolumn}
\usepackage{bm}

\usepackage{comment}
\usepackage[utf8]{inputenc}
\usepackage{soul}
\usepackage{xcolor}

\usepackage{booktabs}

\begin{document}

\preprint{APS/123-QED}

\title{Extended Haloscope Search and Candidate Validation near 1.036\,GHz}


\author{Saebyeok Ahn}
\affiliation{Dark Matter Axion Group, Institute for Basic Science, Daejeon 34126, Republic of Korea}
\affiliation{Center for Axion and Precision Physics Research, Institute for Basic Science (IBS), Daejeon 34051, Republic of Korea}


\author{Boris I. Ivanov}
\affiliation{Dark Matter Axion Group, Institute for Basic Science, Daejeon 34126, Republic of Korea}
\affiliation{Center for Axion and Precision Physics Research, Institute for Basic Science (IBS), Daejeon 34051, Republic of Korea}

\author{Ohjoon Kwon}
\affiliation{Dark Matter Axion Group, Institute for Basic Science, Daejeon 34126, Republic of Korea}
\affiliation{Center for Axion and Precision Physics Research, Institute for Basic Science (IBS), Daejeon 34051, Republic of Korea}

\author{HeeSu Byun}
\affiliation{Center for Axion and Precision Physics Research, Institute for Basic Science (IBS), Daejeon 34051, Republic of Korea}

\author{Arjan F. van Loo}
\affiliation{RIKEN Center for Quantum Computing (RQC), Wako, Saitama 351-0198, Japan}
\affiliation{Department of Applied Physics, Graduate School of Engineering, The University of Tokyo, Bunkyo-ku, Tokyo 113-8656, Japan}

\author{SeongTae Park}
\affiliation{Center for Axion and Precision Physics Research, Institute for Basic Science (IBS), Daejeon 34051, Republic of Korea}

\author{JinMyeong Kim}
\affiliation{Department of Physics, Korea Advanced Institute of Science and Technology (KAIST), Daejeon 34141, Republic of Korea}
\affiliation{Center for Axion and Precision Physics Research, Institute for Basic Science (IBS), Daejeon 34051, Republic of Korea}

\author{Junu Jeong}
\thanks{Present address: Oskar Klein Centre, Department of Physics, Stockholm University, AlbaNova, SE-10691 Stockholm, Sweden}
\affiliation{Center for Axion and Precision Physics Research, Institute for Basic Science (IBS), Daejeon 34051, Republic of Korea}

\author{Soohyung Lee}
\thanks{Present address: Center for Accelerator Research, Korea University, Sejong 30019, Republic of Korea}
\affiliation{Center for Axion and Precision Physics Research, Institute for Basic Science (IBS), Daejeon 34051, Republic of Korea}

\author{Jinsu Kim}
\affiliation{Dark Matter Axion Group, Institute for Basic Science, Daejeon 34126, Republic of Korea}
\affiliation{Center for Axion and Precision Physics Research, Institute for Basic Science (IBS), Daejeon 34051, Republic of Korea}

\author{Çağlar Kutlu}
\thanks{Present address: Zurich Instruments, Technoparkstrasse 1, 8005 Zürich, Switzerland.}
\affiliation{Center for Axion and Precision Physics Research, Institute for Basic Science (IBS), Daejeon 34051, Republic of Korea}

\author{Andrew K. Yi}
\thanks{Present address: SLAC National Accelerator Laboratory, 2575 Sand Hill Road, Menlo Park, California 94025, USA}
\affiliation{Department of Physics, Korea Advanced Institute of Science and Technology (KAIST), Daejeon 34141, Republic of Korea}
\affiliation{Center for Axion and Precision Physics Research, Institute for Basic Science (IBS), Daejeon 34051, Republic of Korea}

\author{Yasunobu Nakamura}
\affiliation{RIKEN Center for Quantum Computing (RQC), Wako, Saitama 351-0198, Japan}
\affiliation{Department of Applied Physics, Graduate School of Engineering, The University of Tokyo, Bunkyo-ku, Tokyo 113-8656, Japan}

\author{Seonjeong Oh}
\affiliation{Dark Matter Axion Group, Institute for Basic Science, Daejeon 34126, Republic of Korea}
\affiliation{Center for Axion and Precision Physics Research, Institute for Basic Science (IBS), Daejeon 34051, Republic of Korea}

\author{Danho Ahn}
\thanks{Present address: INFN-Sezione di Padova, Via Marzolo 8, 35131 Padova, Italy}
\affiliation{Center for Axion and Precision Physics Research, Institute for Basic Science (IBS), Daejeon 34051, Republic of Korea}

\author{SungJae Bae}
\thanks{Present address: RIKEN Center for Quantum Computing (RQC), Wako, Saitama 351–0198, Japan}
\affiliation{Department of Physics, Korea Advanced Institute of Science and Technology (KAIST), Daejeon 34141, Republic of Korea}
\affiliation{Center for Axion and Precision Physics Research, Institute for Basic Science (IBS), Daejeon 34051, Republic of Korea}

\author{Hyoungsoon Choi}
\affiliation{Department of Physics, Korea Advanced Institute of Science and Technology (KAIST), Daejeon 34141, Republic of Korea}

\author{Jihoon Choi}
\thanks{Present address: Korea Astronomy and Space Science Institute, Daejeon 34055, Republic of Korea.}
\affiliation{Center for Axion and Precision Physics Research, Institute for Basic Science (IBS), Daejeon 34051, Republic of Korea}

\author{Yonuk Chong}
\affiliation{SKKU Advancd Institute of Nano Technology (SAINT) and Department of Nano Engineering
Sung Kyun Kwan University (SKKU), Suwon 16419, Republic of Korea}

\author{Woohyun Chung}
\affiliation{Center for Axion and Precision Physics Research, Institute for Basic Science (IBS), Daejeon 34051, Republic of Korea}

\author{Violeta Gkika}
\affiliation{Center for Axion and Precision Physics Research, Institute for Basic Science (IBS), Daejeon 34051, Republic of Korea}

\author{Jihn E. Kim}
\affiliation{Department of Physics, Seoul National University, 1 Gwanak-Ro, Seoul 08826, Republic of Korea}

\author{Younggeun Kim}
\thanks{Present address: Johannes Gutenberg-Universität Mainz, 55122 Mainz, Germany; GSI Helmholtzzentrum für Schwerionenforschung GmbH, 64291 Darmstadt, Germany}
\affiliation{Center for Axion and Precision Physics Research, Institute for Basic Science (IBS), Daejeon 34051, Republic of Korea}

\author{Byeong Rok Ko}
\thanks{Present address: Department of Accelerator Science, Korea University Sejong Campus, 2511 Sejong-ro, Sejong, 30019, Republic of Korea}
\affiliation{Center for Axion and Precision Physics Research, Institute for Basic Science (IBS), Daejeon 34051, Republic of Korea}

\author{Lino Miceli}
\affiliation{Center for Axion and Precision Physics Research, Institute for Basic Science (IBS), Daejeon 34051, Republic of Korea}

\author{Doyu Lee}
\thanks{Present address: Samsung Electronics, Gyeonggi-do 16677, Republic of Korea.}
\affiliation{Center for Axion and Precision Physics Research, Institute for Basic Science (IBS), Daejeon 34051, Republic of Korea}

\author{Jiwon Lee}
\affiliation{Dark Matter Axion Group, Institute for Basic Science, Daejeon 34126, Republic of Korea}
\affiliation{Department of Physics, Korea Advanced Institute of Science and Technology (KAIST), Daejeon 34141, Republic of Korea}
\affiliation{Center for Axion and Precision Physics Research, Institute for Basic Science (IBS), Daejeon 34051, Republic of Korea}

\author{Ki Woong Lee}
\affiliation{Center for Axion and Precision Physics Research, Institute for Basic Science (IBS), Daejeon 34051, Republic of Korea}

\author{MyeongJae Lee}
\thanks{Present address: Department of Physics, Sungkyunkwan University, Suwon 16419, Republic of Korea.}
\affiliation{Center for Axion and Precision Physics Research, Institute for Basic Science (IBS), Daejeon 34051, Republic of Korea}

\author{Andrei Matlashov}
\thanks{Deceased.}
\affiliation{Center for Axion and Precision Physics Research, Institute for Basic Science (IBS), Daejeon 34051, Republic of Korea}

\author{Pallavi Parashar}
\affiliation{Dark Matter Axion Group, Institute for Basic Science, Daejeon 34126, Republic of Korea}
\affiliation{Department of Physics, Korea Advanced Institute of Science and Technology (KAIST), Daejeon 34141, Republic of Korea}
\affiliation{Center for Axion and Precision Physics Research, Institute for Basic Science (IBS), Daejeon 34051, Republic of Korea}

\author{Taehyeon Seong}
\affiliation{Dark Matter Axion Group, Institute for Basic Science, Daejeon 34126, Republic of Korea}
\affiliation{Center for Axion and Precision Physics Research, Institute for Basic Science (IBS), Daejeon 34051, Republic of Korea}

\author{Yun Chang Shin}
\affiliation{Center for Axion and Precision Physics Research, Institute for Basic Science (IBS), Daejeon 34051, Republic of Korea}

\author{Sergey V. Uchaikin}
\affiliation{Dark Matter Axion Group, Institute for Basic Science, Daejeon 34126, Republic of Korea}
\affiliation{Center for Axion and Precision Physics Research, Institute for Basic Science (IBS), Daejeon 34051, Republic of Korea}

\author{Yannis K. Semertzidis}
\affiliation{Center for Axion and Precision Physics Research, Institute for Basic Science (IBS), Daejeon 34051, Republic of Korea}
\affiliation{Department of Physics, Korea Advanced Institute of Science and Technology (KAIST), Daejeon 34141, Republic of Korea}

\author{SungWoo Youn}
\thanks{corresponding author}
\email{swyoun@ibs.re.kr}
\affiliation{Dark Matter Axion Group, Institute for Basic Science, Daejeon 34126, Republic of Korea}
\affiliation{Center for Axion and Precision Physics Research, Institute for Basic Science (IBS), Daejeon 34051, Republic of Korea}

\date{\today}

\begin{abstract}
We report a follow-up axion haloscope search near 1.036 GHz that completes and extends our previous work [Phys. Rev. X \textbf{14}, 031023 (2024)], in which a portion of the HEMT-based data could not be analyzed due to unrecorded experimental information. 
While recovering this dataset, we identified an excess near 1.036 GHz that satisfied our candidate-selection criteria, motivating dedicated validation studies, including independent cross-checks and re-examination with the original apparatus. 
The excess did not persist under these investigations and was not confirmed as an axion dark-matter signal.
We subsequently extended the search over a 20-MHz band surrounding the candidate using a quantum-noise-limited amplifier, achieving sensitivity close to the Dine–Fischler–Srednicki–Zhitnitsky benchmark. 
In the absence of a confirmed signal, we set improved 90\% confidence-level upper limits on the axion–photon coupling over the frequency range 1.026–1.045 GHz. 
This work highlights the importance of robust candidate-validation strategies as haloscope searches approach discovery-level sensitivity.
\end{abstract}

\maketitle

The axion, originally proposed as a solution to the strong charge–parity problem in quantum chromodynamics~\cite{PRL37_8_1976,PRD14_3432_1978,PRD18_2199_1978,PR108_120_1957,PRD15_9_1977,NPA341_269_1980,PRL38_1440_1977,PRL40_223_1978,PRL40_279_1978}, is a well-motivated candidate for cold dark matter~\cite{AA594_A13_2016}. 
Resonant microwave haloscopes search for axion dark matter through axion-to-photon conversion in a strong magnetic field, with the conversion power enhanced by a high-$Q$ cavity.
The expected signal power is given by~\cite{CAPP_scan_rate}
\begin{equation}
P_{\rm sig} = g_{a\gamma\gamma}^2 \, \frac{\rho_a}{m_a} \, B_0^2 V C Q \frac{\beta}{\beta+1},
\end{equation}
where $g_{a\gamma\gamma}$ denotes the axion–photon coupling, $\rho_a$ the local axion dark-matter density, $m_a$ the axion mass, $B_0$ the applied magnetic field strength, $V$ the cavity volume, $C$ the mode-dependent form factor, $Q$ the cavity quality factor, and $\beta$ the antenna coupling.
Benchmark axion models such as Kim–Shifman–Vainshtein–Zakharov~(KSVZ)~\cite{PRL43_103_1979,NPB166_493_1980} and Dine–Fischler–Srednicki–Zhitnitsky~(DFSZ)~\cite{YF31_497_1980,PLB104_199_1981} define the natural sensitivity goals for haloscope experiments.
As advances in cryogenic amplification bring receiver noise toward the quantum limit, haloscope experiments are entering a regime of discovery-level sensitivity. 
In this regime, rigorous candidate identification and validation become essential to distinguish genuine axion signals from instrumental artifacts and environmental radio-frequency interference.

In a recent search reported in Ref.~\cite{CAPP-MAX}, we set stringent limits on the axion–photon coupling over a broad frequency range.
Most of the range was scanned using Josephson parametric amplifiers (JPAs)~\cite{CAPP_jpa_frontier,CAPP_jpa_char}, while certain intervals outside the JPA bandwidth were covered with high-electron-mobility transistor (HEMT) amplification.
A subset of data in the 1.033--1.037\,GHz range could not be analyzed due to missing antenna-coupling measurements and was therefore excluded from the original report.

In this Letter, we complete the missing frequency band to provide continuous exclusion coverage and establish a reference point for subsequent higher-sensitivity searches.
The restored data revealed a feature near 1.036\,GHz that met our candidate-selection criteria, prompting a comprehensive program of validation studies.
We report the outcome of these investigations: independent cross-checks with a separate haloscope system and re-examination using the original apparatus. 
We subsequently extended the search using a JPA receiver and report updated 90\% confidence-level limits on the axion–photon coupling over the frequency range 1.026--1.045 GHz.

\begin{figure}[ht]
    \centering
    \includegraphics[width=0.9\linewidth]{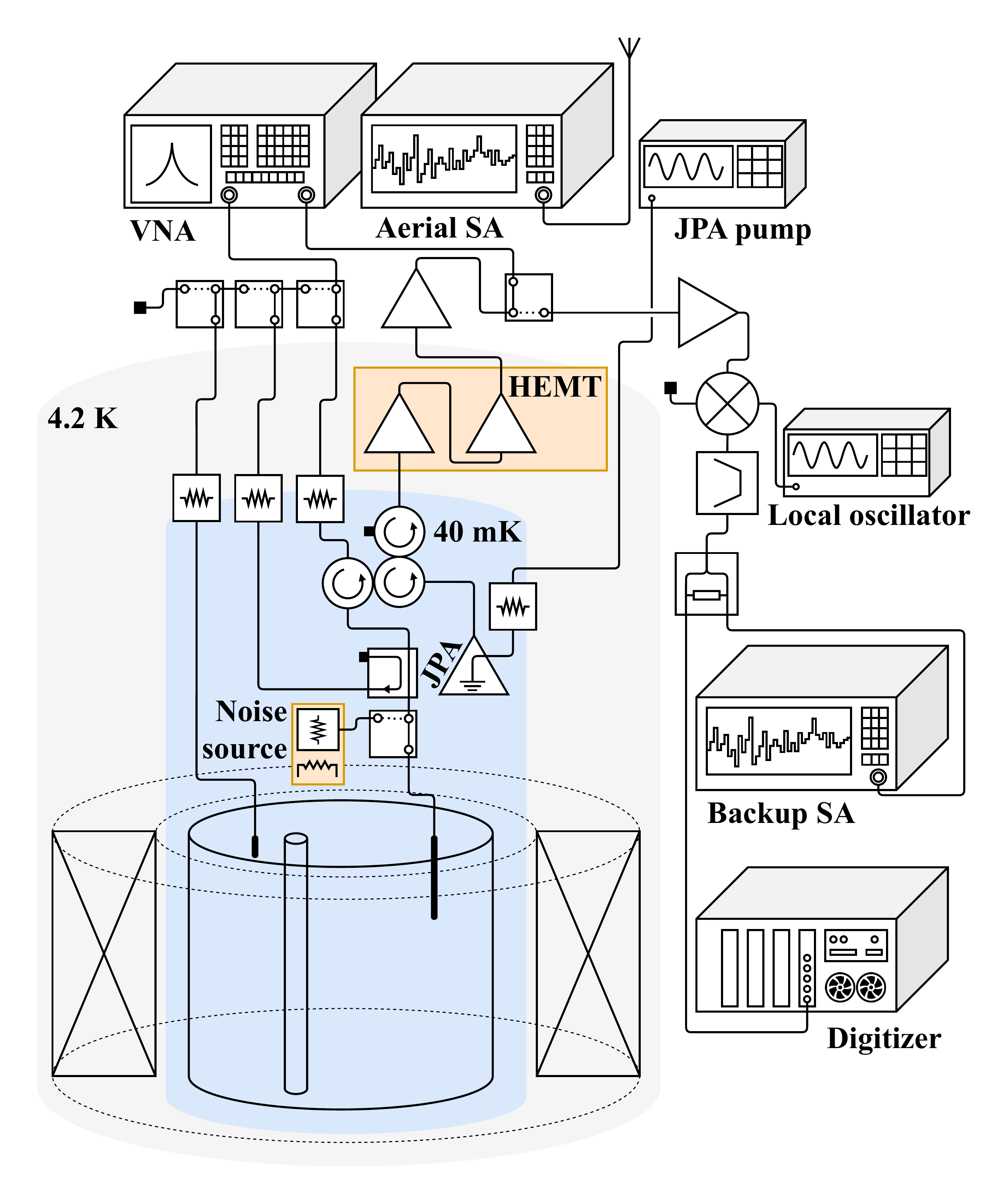}
    \caption{Schematic of the axion haloscope setup, including the microwave cavity in a superconducting magnet and the associated cryogenic and room-temperature receiver chain. }
    \label{fig:schematics}
\end{figure}

The original scan~\cite{CAPP-MAX} was performed with a haloscope, featuring a 12-T superconducting solenoid, a 37-L cylindrical copper cavity and near-quantum-limited JPAs.
The system was maintained below 40\,mK by a dilution refrigerator, substantially suppressing thermal noise from both the cavity and the receiver chain.
The experiment operated in the TM$_{010}$ cavity mode, which maximizes axion–photon conversion power, with frequency tuning achieved via a piezoelectric-actuated copper rod inside the cavity.
The cavity signal was extracted via a strongly coupled coaxial antenna with $\beta\sim1.5$. 
The receiver chain comprised JPAs operated with gains of $\sim$20\,dB with a bandwidth of $\sim$100\,kHz, followed by cryogenic HEMT amplifiers and subsequent room-temperature amplification.
The cavity resonant frequency was set 100-kHz away from the JPA resonance in order to operate the JPA in the phase-insensitive mode.
The amplified signals were down-converted to an intermediate frequency of 10.7\,MHz using an image-rejection mixer and digitized with a resolution bandwidth of 10\,Hz. 
A dedicated aerial antenna system was employed to monitor environmental radio-frequency interference.  
A schematic of the experimental setup is shown in Fig.~\ref{fig:schematics}.

In the original search, near-quantum-limited sensitivity was achieved using multiple JPAs operated in a parallel configuration, in which a single pump tone simultaneously drove individual amplifiers with different frequency bandwidths.
The effective resonant frequency was selected by the pump frequency, enabling an extended search bandwidth while sharing a common circulator and receiver chain.
Although JPAs provided near-quantum-limited amplification over most of the frequency range, frequency intervals outside their bandwidth were necessarily scanned with HEMTs.
In one such dataset spanning 1.033--1.037\,GHz, the antenna coupling was not recorded, and the data were therefore excluded from the previous analysis.
We subsequently recovered the coupling by exploiting the smooth frequency dependence of the loaded quality factor $Q_L$, which was measured across the full range. 
Motivated by this behavior, we assumed the antenna coupling to vary smoothly with frequency and inferred $\beta$ in the missing region by interpolating coupling values measured in adjacent frequency regions. 
A quadratic interpolation was employed to reflect the overall frequency dependence. 
The resulting reconstruction showed negligible boundary discontinuities, and the inferred $\beta$ was insensitive to the interpolation order, with a maximum deviation of $\sim$0.37\%.

\begin{figure}[ht]
    \centering
    \includegraphics[width=\linewidth]{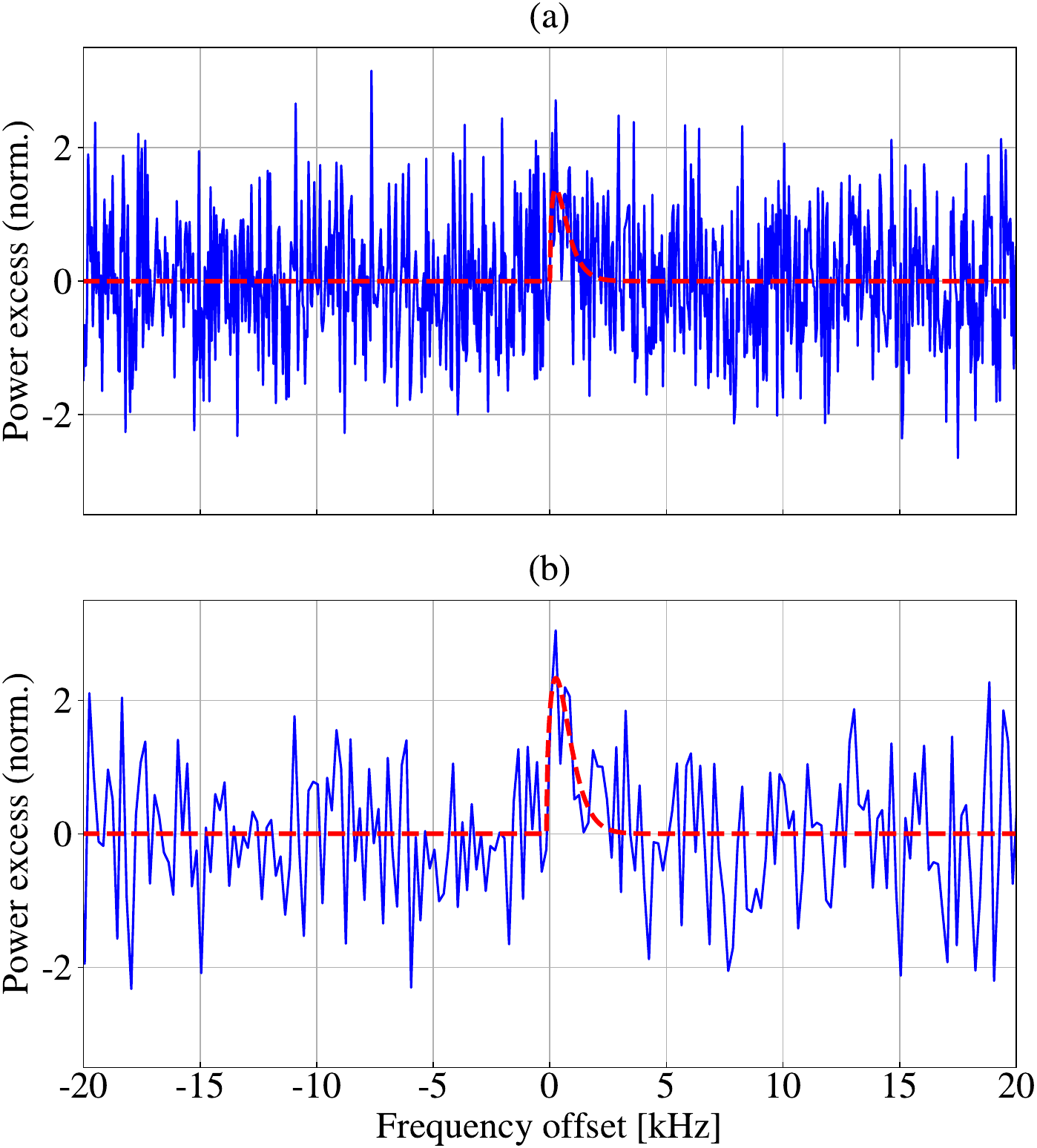}
    \caption{Power spectra centered on the candidate frequency for different frequency-bin widths: (a) 50\,Hz and (b) 200\,Hz. 
    The expected signal lineshape from the standard axion halo model is overlaid as red dashed lines.}
    \label{fig:signal_in_bins} 
\end{figure}

\begin{figure}[ht]
    \centering
    \includegraphics[width=\linewidth]{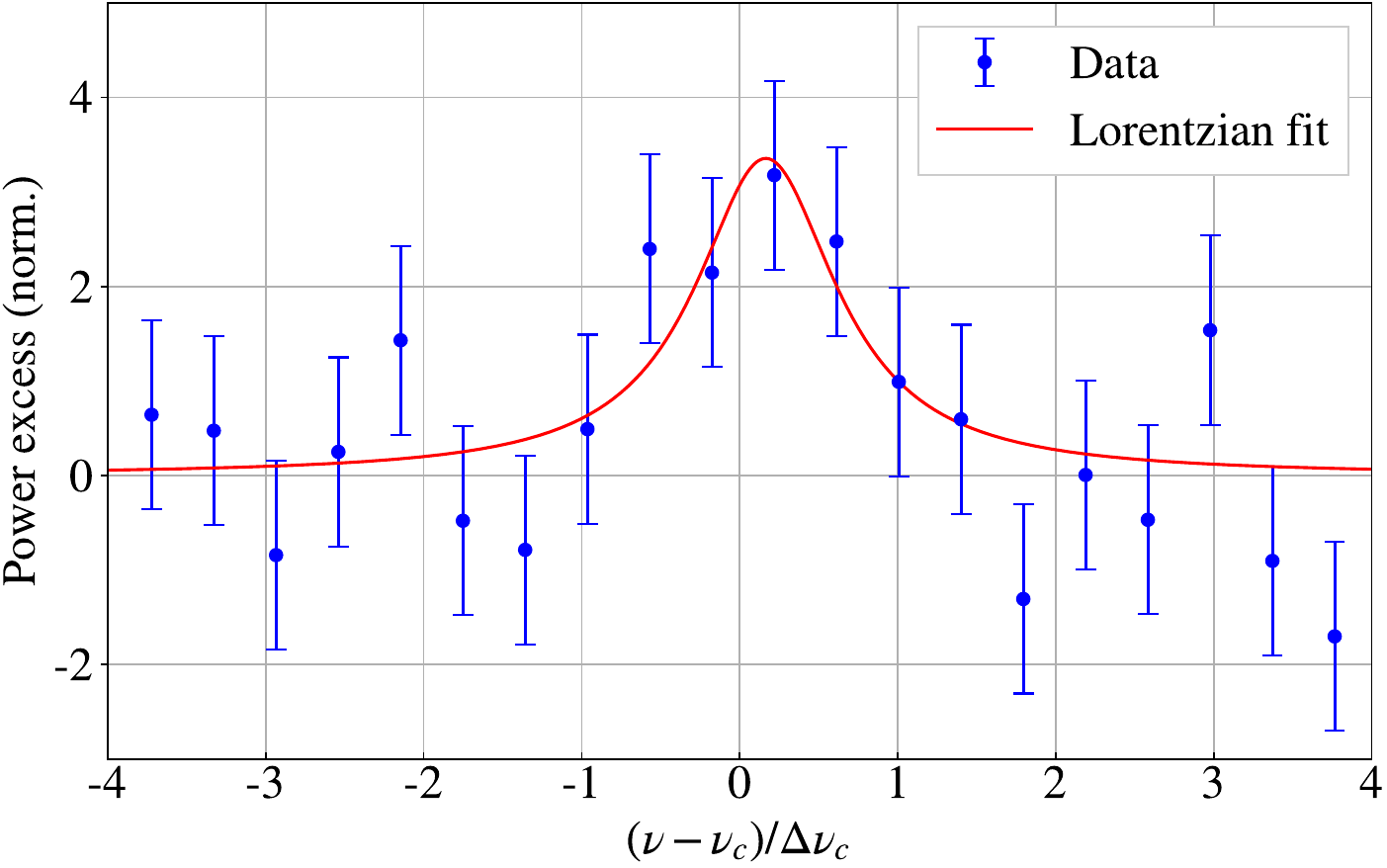}
    \caption{Signal strength as a function of frequency detuning $\nu-\nu_c$, normalized to the cavity bandwidth ($\Delta\nu_c=28.25$\,kHz).
    The error bars indicate statistical uncertainties.
    The data are fitted with a Lorentzian profile whose linewidth is consistent with the measured cavity resonance.}
    \label{fig:cavity_lineshape}
\end{figure}

During the analysis of the recovered dataset, an excess was observed at 1.036315\,GHz ($m_a=4.285857\,\mu$eV) in data acquired on January 23, 2023. 
The excess exhibited a local statistical significance of $5.1\sigma$, which is reduced to a global significance of $3.5\sigma$ after accounting for the look-elsewhere effect. 
The observed spectral profile was consistent with that expected from virialized axion dark matter~\cite{PhysRevD.42.3572}. 
The corresponding power spectra are shown in Fig.~\ref{fig:signal_in_bins}. 
The signal strength as a function of frequency tuning is consistent with the cavity resonance response, as shown in Fig.~\ref{fig:cavity_lineshape}.
The inferred signal strength corresponded to roughly $1.3$ times the KSVZ axion--photon coupling, assuming a local axion density of $0.45\,\mathrm{GeV/cm^{3}}$~\cite{10.1093/mnras/stae034, Lim_2025}. 
Given its statistical significance and characteristic spectral profile, the excess was treated as an axion candidate and subjected to follow-up investigations.

A cross-check was performed using an independent haloscope system, previously utilized for an early axion search~\cite{CAPP-1}. 
The apparatus consists of an 8-T superconducting magnet with a 165-mm bore and a cryogen-free dilution refrigerator operating below 40\,mK. 
For this study, a dedicated cylindrical copper cavity with internal dimensions of $\O 134\,{\rm mm} \times 246\,{\rm mm}$ was fabricated.
Its natural TM$_{010}$ resonance at 1.713\,GHz was tuned down to the candidate frequency by inserting a $\O 30$-mm alumina rod.
This modification resulted in a reduced form factor of $C_{010} \approx 0.12$, as estimated from electromagnetic numerical eigenmode computations.
Quantum-limited amplification was achieved using a single flux-driven JPA optimized for the target frequency. 
An initial dataset acquired in December 2023, taken with a cryogenic HEMT amplifier exhibiting elevated system noise ($T_{\rm sys}\sim500$\,mK), showed a transient excess at the candidate frequency with an observed SNR of $\sim$3.7, prompting further integration to assess its persistence; however, the excess did not reappear with increased exposure.
Data collected in a subsequent run in January 2024, after replacement of a cryogenic HEMT amplifier that reduced the system noise to $\sim$200\,mK, revealed no statistically significant excess, failing to confirm the candidate.
 
Following the independent cross-check, we re-examined the candidate using the original apparatus reconfigured with the same single JPA used in the cross-check experiment.
This configuration enabled a direct test of whether the previously observed excess originated from systematic effects associated with the amplification chain or other setup-specific artifacts.
A follow-up scan conducted between April and June 2024, comprising several segmented data-taking periods due to system instability that required repeated helium recondensing procedures, found no persistent signal with a spectral profile or temporal stability consistent with virialized axion dark matter.
We therefore find no evidence supporting a dark-matter interpretation of the candidate and conclude that the initially observed excess does not indicate galactic axion dark matter, although its precise origin remains unidentified.
We note that alternative scenarios involving clustered or transient axion dark matter have been proposed in the literature~\cite{Tinyakov_2016, PhysRevD.101.023008}; however, the present data are not sufficient to support such interpretations.

\begin{figure}[ht]
    \centering
    \includegraphics[width=0.95\linewidth]{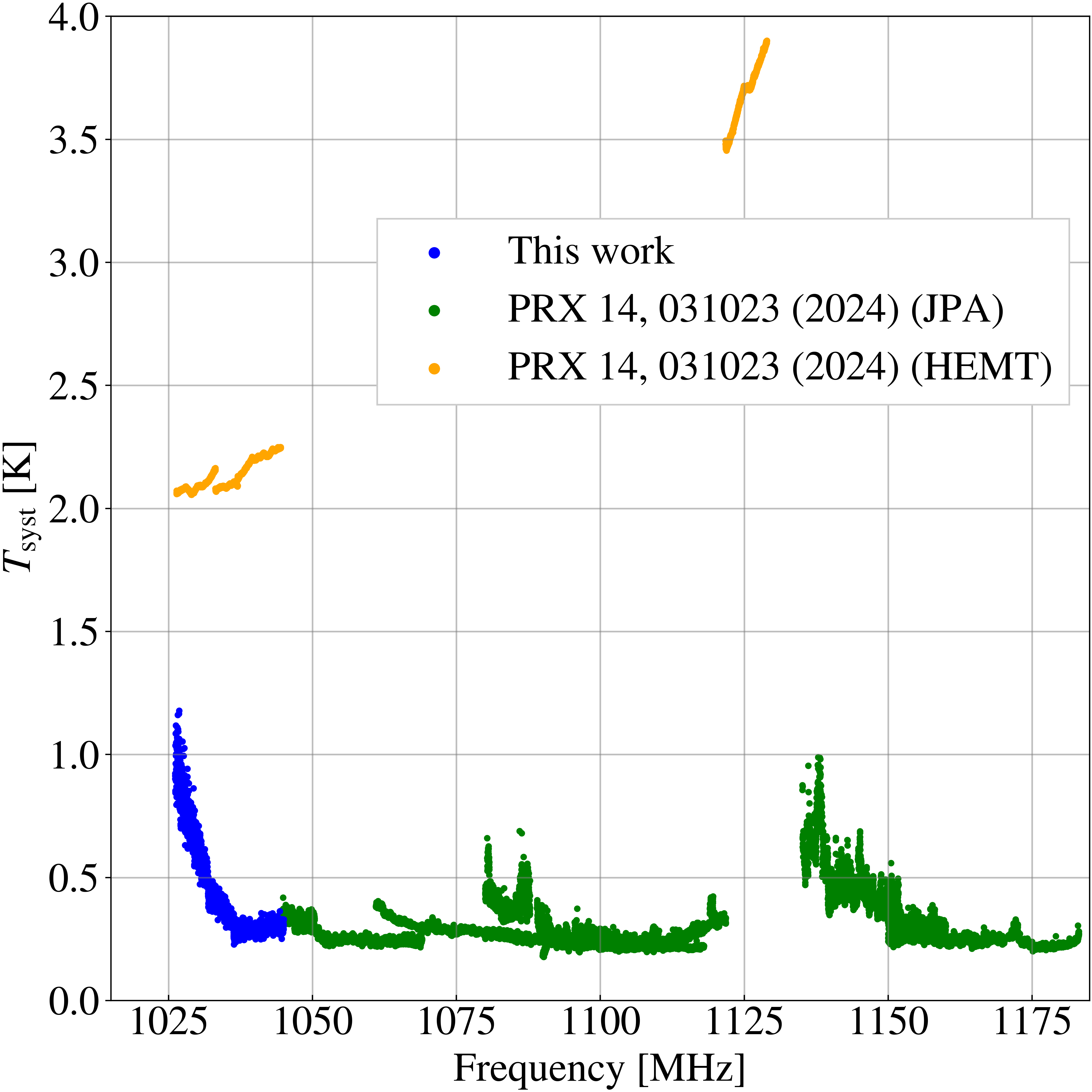}
    \caption{Measured system noise temperature $T_{\rm sys}$ for this scan (blue), compared with the original scans employing JPA-based (green) and HEMT-based (yellow) receiver chains~\cite{CAPP-MAX}. 
    The red dashed line indicates the standard quantum limit.}
    \label{fig:noise}
\end{figure}

To further strengthen this conclusion, we performed a targeted rescan over the $\sim$20-MHz band surrounding the candidate frequency. 
Data was collected over a total of 12 days from June 20 to July 10, 2024, with an interruption due to facility maintenance. 
The analysis procedure followed Ref.~\cite{CAPP-MAX}, consisting of baseline estimation, spectral preprocessing, and statistical combination. 
Baselines extracted from cavity transmission and reflection spectra, JPA noise and gain spectra are used to determine the cavity parameters, expected signal power, and system noise temperature. 
The performance of the baseline removal and normalization was verified by the resulting spectra following standard normal distributions.
Correlations between mirrored frequency bins induced by the JPA were accounted for by optimally combining correlated pairs using their expected SNRs and measured correlation coefficients, yielding an average SNR improvement of 4.1\%. 
The analysis efficiency was quantified by injecting software-synthesized axion signals with a reference SNR of 5 directly into the obtained power spectra, and comparing the injected and recovered SNRs after the full analysis chain. 
Repeating this procedure across the scanned frequency range, we found an average efficiency of $92.7\pm0.9\%$.

Spectra from different cavity tunings were combined using SNR weighting. 
In the resulting spectrum, 146 clusters exceeding $5 \sigma$ were identified; all were found to be inconsistent with axion dark matter signals based on their narrow bandwidths, correlations with aerial antenna data, lack of the expected cavity response, or non-persistence upon rescanning, and were therefore removed together with neighboring bins. 
Frequency bins were further combined within 5-kHz windows using the expected axion signal shape from the standard halo model. 
A total of 16 clusters exceeding the rescan threshold of $3.718\sigma$ underwent follow-up measurements, yielding one persistent candidate, which was subsequently excluded as a galactic axion signal based on its persistence with the magnetic field turned off.
While incompatible with axion dark matter, such behavior is consistent with magnetic-field–independent scenarios, such as dark photon dark matter, and may warrant further investigation.

The dominant source of uncertainty in the analysis arises from the estimation of the system noise. 
The noise temperature was measured at each frequency-tuning step, and its slow variation across the scanned frequency range was removed using a Savitzky–Golay filter~\cite{SGfilter}. 
The residual fluctuations about the mean were taken as the associated uncertainty, yielding a fractional uncertainty of 6.4\%. 
Figure~\ref{fig:noise} compares the system noise measured over the rescanned frequency range with that obtained in the original scan over the full frequency range. 
Other analysis uncertainties, including those associated with the cavity parameters, contribute at the sub-1\% level.

\begin{figure*}[ht]
    \centering
    \includegraphics[width=\linewidth]{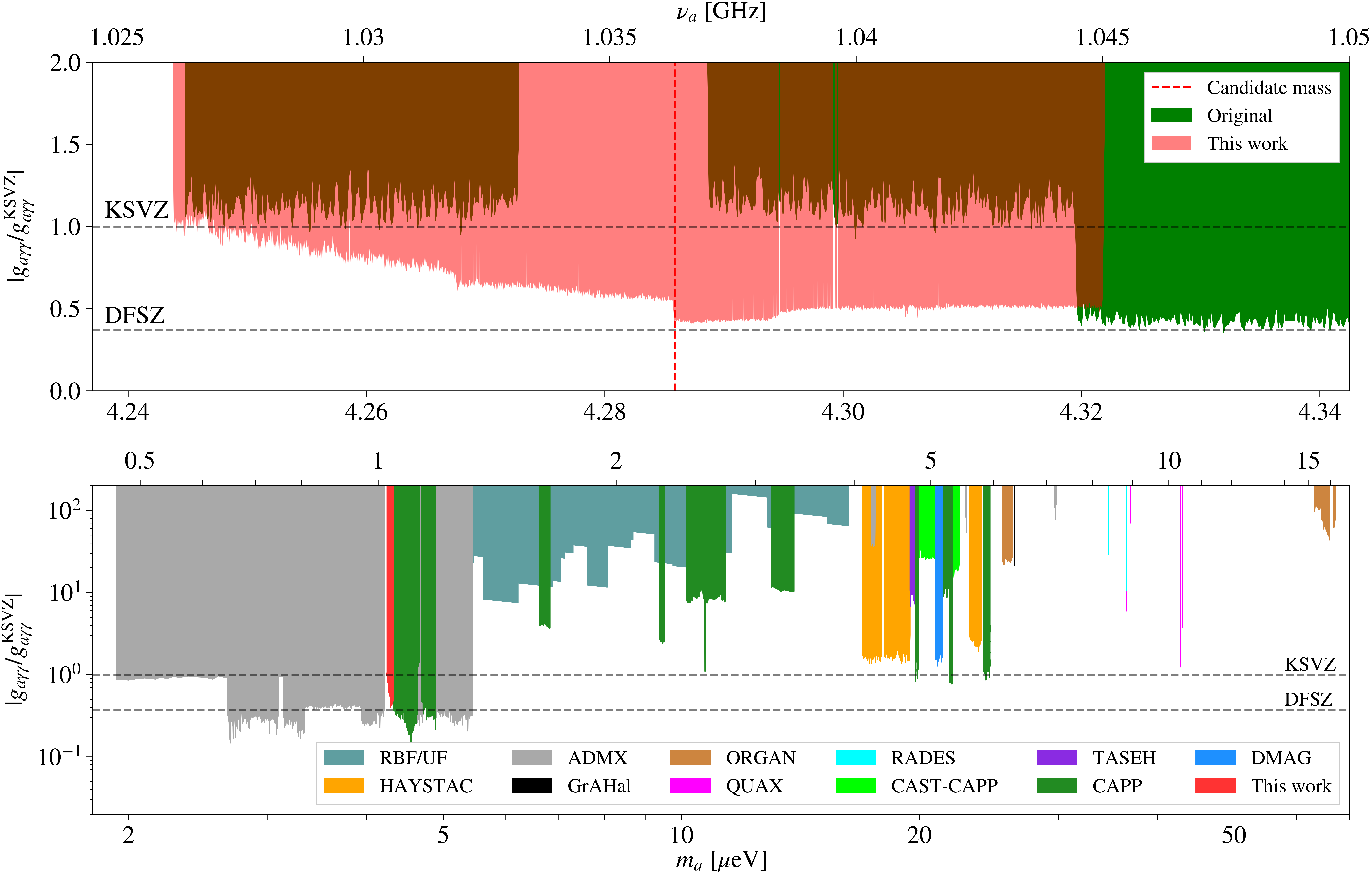}
    \caption{Experimental limits on the axion–photon coupling $g_{a\gamma\gamma}$ as a function of axion mass $m_a$ (frequency $\nu_{a}$).
    The upper panel shows the 90\% confidence-level limits set by this work, assuming a local axion density of $\rho_a=0.45\,{\rm GeV/cm^3}$; the black dashed lines indicate the KSVZ and DFSZ benchmark models. 
    The vertical red dashed line denotes the candidate frequency. 
    The lower panel compares the results of this work with previously reported haloscope limits~\cite{ADMX, ADMX2018, ADMX2019_1_2, ADMX2021, ADMX2024, ADMX2025, ADMX_Sidecar, ADMX_Sidecar_JTWPA, CAPP-1, CAPP-2, CAPP-3, CAPP-4, CAPP-5, CAPP-6, CAPP-7, CAPP-8, CAPP-9, CAPP-MAX, CAPP-8TB-PR2, DMAG-12T, CASTCAPP, GrAHal, HAYSTAC2, HAYSTAC3, HAYSTAC4, ORGAN, QUAX, QUAX2, QUAX3, QUAX4, QUAX5, RADES, RADES2, RBF, UF, TASEH}, with the exclusion data taken from the AxionLimits GitHub repository~\cite{AxionLimits}.}
    \label{fig:exclusion}
\end{figure*}

With no signal consistent with axion dark matter observed, we set 90\% confidence-level upper limits on the axion--photon coupling over a frequency (mass) range 1.026--1.045\,GHz (4.244--4.322\,$\mu$eV), as shown in Fig.~\ref{fig:exclusion}.
Due to the performance of the JPA, the upper portion of this range achieved sensitivity approaching the DFSZ benchmark, while the lower portion reached sensitivity at the KSVZ level.
These results supersede previous constraints and represent among the most stringent haloscope limits reported to date in this range.

In summary, we completed an axion haloscope search by recovering the previously unrecorded frequency region near 1.036\,GHz from Ref.~\cite{CAPP-MAX}.
A candidate excess was identified and subjected to a rigorous program of validation, ultimately failing to confirm as an axion dark matter signal. 
Independent measurements by the ADMX experiment also probed this frequency region with adequate sensitivity and found no indication of an excess signal (G. Rybka, private communication).
We subsequently extended search around the candidate frequency using a quantum-noise-limited JPA receiver, enabling improved 90\% confidence-level upper limits on the axion–photon coupling.
This work demonstrates the robustness and importance of systematic candidate-validation strategies as axion haloscope searches approach discovery-level sensitivity.

\begin{acknowledgments}
This work was supported by the Institute for Basic Science (IBS-R017-D1-2024-a00 and IBS-R040-C1-2025-a00) and by JSPS KAKENHI (Grant No.~JP22H04937).
A. F. van Loo was supported by a JSPS Postdoctoral Fellowship.
J. E. Kim was partially supported by the Korea Academy of Science.
\end{acknowledgments}

\bibliography{main}

@PREAMBLE{
 "\providecommand{\noopsort}[1]{}" 
 # "\providecommand{\singleletter}[1]{#1}%" 
}

@article{PRL40_223_1978,
    title = {{A New Light Boson?}},
    author = {Weinberg, Steven},
    journal = {Phys. Rev. Lett.},
    volume = {40},
    issue = {4},
    pages = {223--226},
    numpages = {0},
    year = {1978},
    month = {Jan},
    publisher = {American Physical Society},
    doi = {10.1103/PhysRevLett.40.223},
    url = {https://link.aps.org/doi/10.1103/PhysRevLett.40.223}
}

@article{PRL40_279_1978,
    title = {{Problem of Strong $P$ and $T$ Invariance in the Presence of Instantons}},
    author = {Wilczek, F.},
    journal = {Phys. Rev. Lett.},
    volume = {40},
    issue = {5},
    pages = {279--282},
    numpages = {0},
    year = {1978},
    month = {Jan},
    publisher = {American Physical Society},
    doi = {10.1103/PhysRevLett.40.279},
    url = {https://link.aps.org/doi/10.1103/PhysRevLett.40.279}
}

@article{PRL38_1440_1977,
    title = {{$\mathrm{CP}$ Conservation in the Presence of Pseudoparticles}},
    author = {Peccei, R. D. and Quinn, Helen R.},
    journal = {Phys. Rev. Lett.},
    volume = {38},
    issue = {25},
    pages = {1440--1443},
    numpages = {0},
    year = {1977},
    month = {Jun},
    publisher = {American Physical Society},
    doi = {10.1103/PhysRevLett.38.1440},
    url = {https://link.aps.org/doi/10.1103/PhysRevLett.38.1440}
}

@article{PRL37_8_1976,
    title = {{Symmetry Breaking through Bell-Jackiw Anomalies}},
    author = {'t Hooft, G.},
    journal = {Phys. Rev. Lett.},
    volume = {37},
    issue = {1},
    pages = {8--11},
    numpages = {0},
    year = {1976},
    month = {Jul},
    publisher = {American Physical Society},
    doi = {10.1103/PhysRevLett.37.8},
    url = {https://link.aps.org/doi/10.1103/PhysRevLett.37.8}
}

@article{PRD14_3432_1978,
    title = {{Computation of the quantum effects due to a four-dimensional pseudoparticle}},
    author = {'t Hooft, G.},
    journal = {Phys. Rev. D},
    volume = {14},
    issue = {12},
    pages = {3432--3450},
    numpages = {0},
    year = {1976},
    month = {Dec},
    publisher = {American Physical Society},
    doi = {10.1103/PhysRevD.14.3432},
    url = {https://link.aps.org/doi/10.1103/PhysRevD.14.3432}
}

@article{PRD18_2199_1978,
    title = {{Erratum: Computation of the quantum effects due to a four-dimensional pseudoparticle}},
    author = {Hooft, G. 't},
    journal = {Phys. Rev. D},
    volume = {18},
    issue = {6},
    pages = {2199--2200},
    numpages = {0},
    year = {1978},
    month = {Sep},
    publisher = {American Physical Society},
    doi = {10.1103/PhysRevD.18.2199.3},
    url = {https://link.aps.org/doi/10.1103/PhysRevD.18.2199.3}
}

@article{PR108_120_1957,
    title = {{Experimental Limit to the Electric Dipole Moment of the Neutron}},
    author = {Smith, J. H. and Purcell, E. M. and Ramsey, N. F.},
    journal = {Phys. Rev.},
    volume = {108},
    issue = {1},
    pages = {120--122},
    numpages = {0},
    year = {1957},
    month = {Oct},
    publisher = {American Physical Society},
    doi = {10.1103/PhysRev.108.120},
    url = {https://link.aps.org/doi/10.1103/PhysRev.108.120}
}

@article{PRD15_9_1977,
    title = {{Search for an electric dipole moment of the neutron}},
    author = {Dress, W. B. and Miller, P. D. and Pendlebury, J. M. and Perrin, Paul and Ramsey, Norman F.},
    journal = {Phys. Rev. D},
    volume = {15},
    issue = {1},
    pages = {9--21},
    numpages = {0},
    year = {1977},
    month = {Jan},
    publisher = {American Physical Society},
    doi = {10.1103/PhysRevD.15.9},
    url = {https://link.aps.org/doi/10.1103/PhysRevD.15.9}
}

@article{NPA341_269_1980,
    title = {{A search for the electric dipole moment of the neutron using ultracold neutrons}},
    journal = {Nuclear Physics A},
    volume = {341},
    number = {2},
    pages = {269-283},
    year = {1980},
    issn = {0375-9474},
    doi = {https://doi.org/10.1016/0375-9474(80)90313-9},
    url = {https://www.sciencedirect.com/science/article/pii/0375947480903139},
    author = {I.S. Altarev and Yu.V. Borisov and A.B. Brandin and A.I. Egorov and V.F. Ezhov and S.N. Ivanov and V.M. Lobashov and V.A. Nazarenko and G.D. Porsev and V.L. Ryabov and A.P. Serebrov and R.R. Taldaev},
}

@ARTICLE{AA594_A13_2016,
    title = {{Planck2015 results: XIII. Cosmological parameters}},
    author      = "Ade, P. A. R. and others",
    collaboration  = "Planck Collaboration",
    year        = "2016",
    journal     = "Astron. Asttophys.",
    volume      = "594",
    pages       = "A13",
    doi         = "10.1051/0004-6361/201525830",
    eprint      = "1502.01589",
    archivePrefix = "arXiv"
}

@article{PRL43_103_1979,
    title = {{Weak-Interaction Singlet and Strong $\mathrm{CP}$ Invariance}},
    author = {Kim, Jihn E.},
    journal = {Phys. Rev. Lett.},
    volume = {43},
    issue = {2},
    pages = {103--107},
    numpages = {0},
    year = {1979},
    month = {Jul},
    publisher = {American Physical Society},
    doi = {10.1103/PhysRevLett.43.103},
    url = {https://link.aps.org/doi/10.1103/PhysRevLett.43.103}
}

@article{NPB166_493_1980,
    title = {{Can confinement ensure natural CP invariance of strong interactions?}},
    journal = {Nuclear Physics B},
    volume = {166},
    number = {3},
    pages = {493-506},
    year = {1980},
    issn = {0550-3213},
    doi = {https://doi.org/10.1016/0550-3213(80)90209-6},
    url = {https://www.sciencedirect.com/science/article/pii/0550321380902096},
    author = {M.A. Shifman and A.I. Vainshtein and V.I. Zakharov}
}

@ARTICLE{YF31_497_1980,
    title = {{On Possible Suppression of the Axion Hadron Interactions}},
    note        = {In Russian},
    author      = "Zhitnitskii, A. R.",
    year        = "1980",
    journal     = "Yad. Fiz.",
    volume      = "31",
    pages       = "497",
}

@article{PLB104_199_1981,
    title = {{A simple solution to the strong CP problem with a harmless axion}},
    journal = {Physics Letters B},
    volume = {104},
    number = {3},
    pages = {199-202},
    year = {1981},
    issn = {0370-2693},
    doi = {https://doi.org/10.1016/0370-2693(81)90590-6},
    url = {https://www.sciencedirect.com/science/article/pii/0370269381905906},
    author = {Michael Dine and Willy Fischler and Mark Srednicki}
}

@ARTICLE{CASTCAPP,
  title = {{Search for Dark Matter axions with CAST-CAPP}},
  author    = {Adair, C M and Altenm{\"u}ller, K and Anastassopoulos, V and
               Arguedas Cuendis, S and Baier, J and Barth, K and Belov, A and
               Bozicevic, D and Br{\"a}uninger, H and Cantatore, G and Caspers,
               F and Castel, J F and {\c C}etin, S A and Chung, W and Choi, H
               and Choi, J and Dafni, T and Davenport, M and Dermenev, A and
               Desch, K and D{\"o}brich, B and Fischer, H and Funk, W and
               Galan, J and Gardikiotis, A and Gninenko, S and Golm, J and
               Hasinoff, M D and Hoffmann, D H H and D{\'\i}ez Ib{\'a}{\~n}ez,
               D and Irastorza, I G and Jakov{\v c}i{\'c}, K and Kaminski, J
               and Karuza, M and Krieger, C and Kutlu, {\c C} and Laki{\'c}, B
               and Laurent, J M and Lee, J and Lee, S and Luz{\'o}n, G and
               Malbrunot, C and Margalejo, C and Maroudas, M and Miceli, L and
               Mirallas, H and Obis, L and {\"O}zbey, A and {\"O}zbozduman, K
               and Pivovaroff, M J and Rosu, M and Ruz, J and Ruiz-Ch{\'o}liz,
               E and Schmidt, S and Schumann, M and Semertzidis, Y K and
               Solanki, S K and Stewart, L and Tsagris, I and Vafeiadis, T and
               Vogel, J K and Vretenar, M and Youn, S and Zioutas, K},
  journal   = {Nat. Commun.},
  publisher = {pringer Science and Business Media LLC},
  volume    = {13},
  number    = {1},
  pages     = {6180},
  month     = {oct},
  year      = {2022},
  copyright = {https://creativecommons.org/licenses/by/4.0},
}

@ARTICLE{HAYSTAC2,
  title     = "A quantum enhanced search for dark matter axions",
  author    = "Backes, K M and Palken, D A and Kenany, S Al and Brubaker, B M
               and Cahn, S B and Droster, A and Hilton, Gene C and Ghosh,
               Sumita and Jackson, H and Lamoreaux, S K and Leder, A F and
               Lehnert, K W and Lewis, S M and Malnou, M and Maruyama, R H and
               Rapidis, N M and Simanovskaia, M and Singh, Sukhman and Speller,
               D H and Urdinaran, I and Vale, Leila R and van Assendelft, E C
               and van Bibber, K and Wang, H",
  journal   = "Nature",
  publisher = "Springer Science and Business Media LLC",
  volume    =  590,
  number    =  7845,
  pages     = "238--242",
  month     =  feb,
  year      =  2021,
}

@ARTICLE{HAYSTAC3,
  title = {{New results from HAYSTAC's phase II operation with a squeezed state receiver}},
  author = {Jewell, M. J. and Leder, A. F. and Backes, K. M. and Bai, Xiran and van Bibber, K. and Brubaker, B. M. and Cahn, S. B. and Droster, A. and Esmat, Maryam H. and Ghosh, Sumita and Graham, Eleanor and Hilton, Gene C. and Jackson, H. and Laffan, Claire and Lamoreaux, S. K. and Lehnert, K. W. and Lewis, S. M. and Malnou, M. and Maruyama, R. H. and Palken, D. A. and Rapidis, N. M. and Ruddy, E. P. and Simanovskaia, M. and Singh, Sukhman and Speller, D. H. and Vale, Leila R. and Wang, H. and Zhu, Yuqi},
  collaboration = {HAYSTAC Collaboration},
  journal = {Phys. Rev. D},
  volume = {107},
  issue = {7},
  pages = {072007},
  numpages = {20},
  year = {2023},
  month = {Apr},
  publisher = {American Physical Society},
  doi = {10.1103/PhysRevD.107.072007},
  url = {https://link.aps.org/doi/10.1103/PhysRevD.107.072007}
}

@ARTICLE{HAYSTAC4,
  title = {{Dark Matter Axion Search with HAYSTAC Phase II}},
  author = {Bai, Xiran and Jewell, M. J. and Echevers, J. and van Bibber, K. and Droster, A. and Esmat, Maryam H. and Ghosh, Sumita and Graham, Eleanor and Jackson, H. and Laffan, Claire and Lamoreaux, S. K. and Leder, A. F. and Lehnert, K. W. and Lewis, S. M. and Maruyama, R. H. and Nath, R. D. and Rapidis, N. M. and Ruddy, E. P. and Silva-Feaver, M. and Simanovskaia, M. and Singh, Sukhman and Speller, D. H. and Zacarias, Sabrina and Zhu, Yuqi},
  collaboration = {HAYSTAC Collaboration},
  journal = {Phys. Rev. Lett.},
  volume = {134},
  issue = {15},
  pages = {151006},
  numpages = {8},
  year = {2025},
  month = {Apr},
  publisher = {American Physical Society},
  doi = {10.1103/PhysRevLett.134.151006},
  url = {https://link.aps.org/doi/10.1103/PhysRevLett.134.151006}
}

@article{ORGAN,
  title = {{Near-quantum-limited axion dark matter search with the ORGAN experiment around $26\text{ }\text{ }\mathrm{\ensuremath{\mu}}\mathrm{eV}$}},
  author = {Quiskamp, Aaron P. and Flower, Graeme R. and Samuels, Steven and McAllister, Ben T. and Altin, Paul and Ivanov, Eugene N. and Goryachev, Maxim and Tobar, Michael E.},
  journal = {Phys. Rev. D},
  volume = {111},
  issue = {9},
  pages = {095007},
  numpages = {8},
  year = {2025},
  month = {May},
  publisher = {American Physical Society},
  doi = {10.1103/PhysRevD.111.095007},
  url = {https://link.aps.org/doi/10.1103/PhysRevD.111.095007}
}

@misc{AxionLimits,
    author       = {Ciaran O'Hare},
    title = {{cajohare/AxionLimits: AxionLimits}},
    month        = jul,
    year         = 2020,
    publisher    = {Zenodo},
    version      = {v1.0},
    doi          = {10.5281/zenodo.3932430},
    howpublished = {\url{https://cajohare.github.io/AxionLimits/}}
}

@article{SGfilter,
    author = {Savitzky, Abraham. and Golay, M. J. E.},
    title = {{Smoothing and Differentiation of Data by Simplified Least Squares Procedures.}},
    journal = {Analytical Chemistry},
    volume = {36},
    number = {8},
    pages = {1627-1639},
    year = {1964},
    doi = {10.1021/ac60214a047},
    URL = {https://doi.org/10.1021/ac60214a047},
    eprint = {https://doi.org/10.1021/ac60214a047}
}

@article{ADMX,
  title = {{SQUID-Based Microwave Cavity Search for Dark-Matter Axions}},
  author = {Asztalos, S. J. and Carosi, G. and Hagmann, C. and Kinion, D. and van Bibber, K. and Hotz, M. and Rosenberg, L. J and Rybka, G. and Hoskins, J. and Hwang, J. and Sikivie, P. and Tanner, D. B. and Bradley, R. and Clarke, J.},
  journal = {Phys. Rev. Lett.},
  volume = {104},
  issue = {4},
  pages = {041301},
  numpages = {4},
  year = {2010},
  month = {Jan},
  publisher = {American Physical Society},
  doi = {10.1103/PhysRevLett.104.041301},
  url = {https://link.aps.org/doi/10.1103/PhysRevLett.104.041301}
}

@article{ADMX2018,
  title = {{Search for Invisible Axion Dark Matter with the Axion Dark Matter Experiment}},
  author = {Du, N. and Force, N. and Khatiwada, R. and Lentz, E. and Ottens, R. and Rosenberg, L. J and Rybka, G. and Carosi, G. and Woollett, N. and Bowring, D. and Chou, A. S. and Sonnenschein, A. and Wester, W. and Boutan, C. and Oblath, N. S. and Bradley, R. and Daw, E. J. and Dixit, A. V. and Clarke, J. and O'Kelley, S. R. and Crisosto, N. and Gleason, J. R. and Jois, S. and Sikivie, P. and Stern, I. and Sullivan, N. S. and Tanner, D. B and Hilton, G. C.},
  collaboration = {ADMX Collaboration},
  journal = {Phys. Rev. Lett.},
  volume = {120},
  issue = {15},
  pages = {151301},
  numpages = {5},
  year = {2018},
  month = {Apr},
  publisher = {American Physical Society},
  doi = {10.1103/PhysRevLett.120.151301},
  url = {https://link.aps.org/doi/10.1103/PhysRevLett.120.151301}
}

@article{ADMX2019_1_2,
  title = {{Extended Search for the Invisible Axion with the Axion Dark Matter Experiment}},
  author = {Braine, T. and Cervantes, R. and Crisosto, N. and Du, N. and Kimes, S. and Rosenberg, L. J. and Rybka, G. and Yang, J. and Bowring, D. and Chou, A. S. and Khatiwada, R. and Sonnenschein, A. and Wester, W. and Carosi, G. and Woollett, N. and Duffy, L. D. and Bradley, R. and Boutan, C. and Jones, M. and LaRoque, B. H. and Oblath, N. S. and Taubman, M. S. and Clarke, J. and Dove, A. and Eddins, A. and O'Kelley, S. R. and Nawaz, S. and Siddiqi, I. and Stevenson, N. and Agrawal, A. and Dixit, A. V. and Gleason, J. R. and Jois, S. and Sikivie, P. and Solomon, J. A. and Sullivan, N. S. and Tanner, D. B. and Lentz, E. and Daw, E. J. and Buckley, J. H. and Harrington, P. M. and Henriksen, E. A. and Murch, K. W.},
  collaboration = {ADMX Collaboration},
  journal = {Phys. Rev. Lett.},
  volume = {124},
  issue = {10},
  pages = {101303},
  numpages = {6},
  year = {2020},
  month = {Mar},
  publisher = {American Physical Society},
  doi = {10.1103/PhysRevLett.124.101303},
  url = {https://link.aps.org/doi/10.1103/PhysRevLett.124.101303}
}

@article{ADMX2021,
  title = {{Search for Invisible Axion Dark Matter in the $3.3-4.2\text{ }\text{ }\ensuremath{\mu}\mathrm{eV}$ Mass Range}},
  author = {Bartram, C. and Braine, T. and Burns, E. and Cervantes, R. and Crisosto, N. and Du, N. and Korandla, H. and Leum, G. and Mohapatra, P. and Nitta, T. and Rosenberg, L. J and Rybka, G. and Yang, J. and Clarke, John and Siddiqi, I. and Agrawal, A. and Dixit, A. V. and Awida, M. H. and Chou, A. S. and Hollister, M. and Knirck, S. and Sonnenschein, A. and Wester, W. and Gleason, J. R. and Hipp, A. T. and Jois, S. and Sikivie, P. and Sullivan, N. S. and Tanner, D. B. and Lentz, E. and Khatiwada, R. and Carosi, G. and Robertson, N. and Woollett, N. and Duffy, L. D. and Boutan, C. and Jones, M. and LaRoque, B. H. and Oblath, N. S. and Taubman, M. S. and Daw, E. J. and Perry, M. G. and Buckley, J. H. and Gaikwad, C. and Hoffman, J. and Murch, K. W. and Goryachev, M. and McAllister, B. T. and Quiskamp, A. and Thomson, C. and Tobar, M. E.},
  collaboration = {ADMX Collaboration},
  journal = {Phys. Rev. Lett.},
  volume = {127},
  issue = {26},
  pages = {261803},
  numpages = {6},
  year = {2021},
  month = {Dec},
  publisher = {American Physical Society},
  doi = {10.1103/PhysRevLett.127.261803},
  url = {https://link.aps.org/doi/10.1103/PhysRevLett.127.261803}
}

@article{ADMX2024,
   title = {{ADMX Axion Dark Matter Bounds around 3.3 $\ensuremath{\mu}\mathrm{eV}$
 with Dine-Fischler-Srednicki-Zhitnitsky Discovery Ability}},
   volume={134},
   ISSN={1079-7114},
   url={http://dx.doi.org/10.1103/PhysRevLett.134.111002},
   DOI={10.1103/physrevlett.134.111002},
   number={11},
   journal={Physical Review Letters},
   publisher={American Physical Society (APS)},
   author={Goodman, C. and Guzzetti, M. and Hanretty, C. and Rosenberg, L. J. and Rybka, G. and Sinnis, J. and Zhang, D. and Clarke, John and Siddiqi, I. and Chou, A. S. and Hollister, M. and Knirck, S. and Sonnenschein, A. and Caligiure, T. J. and Gleason, J. R. and Hipp, A. T. and Sikivie, P. and Solano, M. E. and Sullivan, N. S. and Tanner, D. B. and Khatiwada, R. and Carosi, G. and Cisneros, C. and Du, N. and Robertson, N. and Woollett, N. and Duffy, L. D. and Boutan, C. and Braine, T. and Lentz, E. and Oblath, N. S. and Taubman, M. S. and Daw, E. J. and Mostyn, C. and Perry, M. G. and Bartram, C. and Dyson, T. A. and Ruppert, S. and Withers, M. O. and Kuo, C. L. and McAllister, B. T. and Buckley, J. H. and Gaikwad, C. and Hoffman, J. and Murch, K. and Goryachev, M. and Hartman, E. and Quiskamp, A. and Tobar, M. E.},
   year={2025},
   month=mar }

@misc{ADMX2025,
      title = {{Search for Axion Dark Matter from 1.1 to 1.3 GHz with ADMX}}, 
      author={ADMX Collaboration and G. Carosi and C. Cisneros and N. Du and S. Durham and N. Robertson and C. Goodman and M. Guzzetti and C. Hanretty and K. Enzian and L. J Rosenberg and G. Rybka and J. Sinnis and D. Zhang and John Clarke and I. Siddiqi and A. S. Chou and M. Hollister and A. Sonnenschein and S. Knirck and T. J. Caligiure and J. R. Gleason and A. T. Hipp and P. Sikivie and M. E. Solano and N. S. Sullivan and D. B. Tanner and R. Khatiwada and L. D. Duffy and C. Boutan and T. Braine and E. Lentz and N. S. Oblath and M. S. Taubman and E. J. Daw and C. Mostyn and M. G. Perry and C. Bartram and J. Laurel and A. Yi and T. A. Dyson and S. Ruppert and M. O. Withers and C. L. Kuo and B. T. McAllister and J. H. Buckley and C. Gaikwad and J. Hoffman and K. Murch and M. Goryachev and E. Hartman and A. Quiskamp and M. E. Tobar},
      year={2025},
      eprint={2504.07279},
      archivePrefix={arXiv},
      primaryClass={hep-ex},
      url={https://arxiv.org/abs/2504.07279}, 
}

@article{ADMX_Sidecar,
  title = {{Piezoelectrically Tuned Multimode Cavity Search for Axion Dark Matter}},
  author = {Boutan, C. and Jones, M. and LaRoque, B. H. and Oblath, N. S. and Cervantes, R. and Du, N. and Force, N. and Kimes, S. and Ottens, R. and Rosenberg, L. J. and Rybka, G. and Yang, J. and Carosi, G. and Woollett, N. and Bowring, D. and Chou, A. S. and Khatiwada, R. and Sonnenschein, A. and Wester, W. and Bradley, R. and Daw, E. J. and Agrawal, A. and Dixit, A. V. and Clarke, J. and O'Kelley, S. R. and Crisosto, N. and Gleason, J. R. and Jois, S. and Sikivie, P. and Stern, I. and Sullivan, N. S. and Tanner, D. B. and Harrington, P. M. and Lentz, E.},
  collaboration = {ADMX Collaboration},
  journal = {Phys. Rev. Lett.},
  volume = {121},
  issue = {26},
  pages = {261302},
  numpages = {7},
  year = {2018},
  month = {Dec},
  publisher = {American Physical Society},
  doi = {10.1103/PhysRevLett.121.261302},
  url = {https://link.aps.org/doi/10.1103/PhysRevLett.121.261302}
}

@article{ADMX_Sidecar_JTWPA,
    title = {{Dark matter axion search using a Josephson Traveling wave parametric amplifier}},
    author = {Bartram, C. and Braine, T. and Cervantes, R. and Crisosto, N. and Du, N. and Leum, G. and Mohapatra, P. and Nitta, T. and Rosenberg, L. J. and Rybka, G. and Yang, J. and Clarke, John and Siddiqi, I. and Agrawal, A. and Dixit, A. V. and Awida, M. H. and Chou, A. S. and Hollister, M. and Knirck, S. and Sonnenschein, A. and Wester, W. and Gleason, J. R. and Hipp, A. T. and Jois, S. and Sikivie, P. and Sullivan, N. S. and Tanner, D. B. and Lentz, E. and Khatiwada, R. and Carosi, G. and Cisneros, C. and Robertson, N. and Woollett, N. and Duffy, L. D. and Boutan, C. and Jones, M. and LaRoque, B. H. and Oblath, N. S. and Taubman, M. S. and Daw, E. J. and Perry, M. G. and Buckley, J. H. and Gaikwad, C. and Hoffman, J. and Murch, K. and Goryachev, M. and McAllister, B. T. and Quiskamp, A. and Thomson, C. and Tobar, M. E. and Bolkhovsky, V. and Calusine, G. and Oliver, W. and Serniak, K.},
    journal = {Review of Scientific Instruments},
    volume = {94},
    number = {4},
    pages = {044703},
    year = {2023},
    month = {04},
    issn = {0034-6748},
    doi = {10.1063/5.0122907},
    url = {https://doi.org/10.1063/5.0122907}
}

@article{CAPP-1,
  title = {{Axion Dark Matter Search around $6.7\text{ }\text{ }\ensuremath{\mu}\mathrm{eV}$}},
  author = {Lee, S. and Ahn, S. and Choi, J. and Ko, B. R. and Semertzidis, Y. K.},
  journal = {Phys. Rev. Lett.},
  volume = {124},
  issue = {10},
  pages = {101802},
  numpages = {5},
  year = {2020},
  month = {Mar},
  publisher = {American Physical Society},
  doi = {10.1103/PhysRevLett.124.101802},
  url = {https://link.aps.org/doi/10.1103/PhysRevLett.124.101802}
}

@article{CAPP-2,
  title = {{Search for Invisible Axion Dark Matter with a Multiple-Cell Haloscope}},
  author = {Jeong, Junu and Youn, SungWoo and Bae, Sungjae and Kim, Jihngeun and Seong, Taehyeon and Kim, Jihn E. and Semertzidis, Yannis K.},
  journal = {Phys. Rev. Lett.},
  volume = {125},
  issue = {22},
  pages = {221302},
  numpages = {6},
  year = {2020},
  month = {Nov},
  publisher = {American Physical Society},
  doi = {10.1103/PhysRevLett.125.221302},
  url = {https://link.aps.org/doi/10.1103/PhysRevLett.125.221302}
}

@article{CAPP-3,
  title = {{First Results from an Axion Haloscope at CAPP around $10.7\text{ }\text{ }\ensuremath{\mu}\mathrm{eV}$}},
  author = {Kwon, Ohjoon and Lee, Doyu and Chung, Woohyun and Ahn, Danho and Byun, HeeSu and Caspers, Fritz and Choi, Hyoungsoon and Choi, Jihoon and Chong, Yonuk and Jeong, Hoyong and Jeong, Junu and Kim, Jihn E. and Kim, Jinsu and Kutlu, \ifmmode \mbox{\c{C}}\else \c{C}\fi{}a\ifmmode \breve{g}\else \u{g}\fi{}lar and Lee, Jihnhwan and Lee, MyeongJae and Lee, Soohyung and Matlashov, Andrei and Oh, Seonjeong and Park, Seongtae and Uchaikin, Sergey and Youn, SungWoo and Semertzidis, Yannis K.},
  journal = {Phys. Rev. Lett.},
  volume = {126},
  issue = {19},
  pages = {191802},
  numpages = {6},
  year = {2021},
  month = {May},
  publisher = {American Physical Society},
  doi = {10.1103/PhysRevLett.126.191802},
  url = {https://link.aps.org/doi/10.1103/PhysRevLett.126.191802}
}

@article{CAPP-4,
  title = {{Searching for Invisible Axion Dark Matter with an 18 T Magnet Haloscope}},
  author = {Lee, Youngjae and Yang, Byeongsu and Yoon, Hojin and Ahn, Moohyun and Park, Heejun and Min, Byeonghun and Kim, DongLak and Yoo, Jonghee},
  journal = {Phys. Rev. Lett.},
  volume = {128},
  issue = {24},
  pages = {241805},
  numpages = {7},
  year = {2022},
  month = {Jun},
  publisher = {American Physical Society},
  doi = {10.1103/PhysRevLett.128.241805},
  url = {https://link.aps.org/doi/10.1103/PhysRevLett.128.241805}
}

@article{CAPP-5,
  title = {{Near-Quantum-Noise Axion Dark Matter Search at CAPP around $9.5\text{ }\text{ }\mathrm{\ensuremath{\mu}}\mathrm{eV}$}},
  author = {Kim, Jinsu and Kwon, Ohjoon and Kutlu, \ifmmode \mbox{\c{C}}\else \c{C}\fi{}a\ifmmode \breve{g}\else \u{g}\fi{}lar and Chung, Woohyun and Matlashov, Andrei and Uchaikin, Sergey and van Loo, Arjan Ferdinand and Nakamura, Yasunobu and Oh, Seonjeong and Byun, HeeSu and Ahn, Danho and Semertzidis, Yannis K.},
  journal = {Phys. Rev. Lett.},
  volume = {130},
  issue = {9},
  pages = {091602},
  numpages = {6},
  year = {2023},
  month = {Feb},
  publisher = {American Physical Society},
  doi = {10.1103/PhysRevLett.130.091602},
  url = {https://link.aps.org/doi/10.1103/PhysRevLett.130.091602}
}

@article{CAPP-6,
  title = {{Axion Dark Matter Search around $4.55\text{ }\text{ }\mathrm{\ensuremath{\mu}}\mathrm{eV}$ with Dine-Fischler-Srednicki-Zhitnitskii Sensitivity}},
  author = {Yi, Andrew K. and Ahn, Saebyeok and Kutlu, \ifmmode \mbox{\c{C}}\else \c{C}\fi{}a\ifmmode \breve{g}\else \u{g}\fi{}lar and Kim, JinMyeong and Ko, Byeong Rok and Ivanov, Boris I. and Byun, HeeSu and van Loo, Arjan F. and Park, SeongTae and Jeong, Junu and Kwon, Ohjoon and Nakamura, Yasunobu and Uchaikin, Sergey V. and Choi, Jihoon and Lee, Soohyung and Lee, MyeongJae and Shin, Yun Chang and Kim, Jinsu and Lee, Doyu and Ahn, Danho and Bae, SungJae and Lee, Jiwon and Kim, Younggeun and Gkika, Violeta and Lee, Ki Woong and Oh, Seonjeong and Seong, Taehyeon and Kim, DongMin and Chung, Woohyun and Matlashov, Andrei and Youn, SungWoo and Semertzidis, Yannis K.},
  journal = {Phys. Rev. Lett.},
  volume = {130},
  issue = {7},
  pages = {071002},
  numpages = {7},
  year = {2023},
  month = {Feb},
  publisher = {American Physical Society},
  doi = {10.1103/PhysRevLett.130.071002},
  url = {https://link.aps.org/doi/10.1103/PhysRevLett.130.071002}
}

@article{CAPP-7,
  title = {{Extended Axion Dark Matter Search Using the CAPP18T Haloscope}},
  author = {Yang, Byeongsu and Yoon, Hojin and Ahn, Moohyun and Lee, Youngjae and Yoo, Jonghee},
  journal = {Phys. Rev. Lett.},
  volume = {131},
  issue = {8},
  pages = {081801},
  numpages = {6},
  year = {2023},
  month = {Aug},
  publisher = {American Physical Society},
  doi = {10.1103/PhysRevLett.131.081801},
  url = {https://link.aps.org/doi/10.1103/PhysRevLett.131.081801}
}

@article{CAPP-8,
  title = {{Experimental Search for Invisible Dark Matter Axions around $22\text{ }\text{ }\mathrm{\ensuremath{\mu}}\mathrm{eV}$}},
  author = {Kim, Younggeun and Jeong, Junu and Youn, SungWoo and Bae, Sungjae and Lee, Kiwoong and van Loo, Arjan F. and Nakamura, Yasunobu and Oh, Seonjeong and Seong, Taehyeon and Uchaikin, Sergey and Kim, Jihn E. and Semertzidis, Yannis K.},
  journal = {Phys. Rev. Lett.},
  volume = {133},
  issue = {5},
  pages = {051802},
  numpages = {6},
  year = {2024},
  month = {Jul},
  publisher = {American Physical Society},
  doi = {10.1103/PhysRevLett.133.051802},
  url = {https://link.aps.org/doi/10.1103/PhysRevLett.133.051802}
}

@article{CAPP-9,
  title = {{Search for Dark Matter Axions with Tunable ${\mathrm{TM}}_{020}$ Mode}},
  author = {Bae, Sungjae and Jeong, Junu and Kim, Younggeun and Youn, SungWoo and Park, Heejun and Seong, Taehyeon and Oh, Seongjeong and Semertzidis, Yannis K.},
  journal = {Phys. Rev. Lett.},
  volume = {133},
  issue = {21},
  pages = {211803},
  numpages = {6},
  year = {2024},
  month = {Nov},
  publisher = {American Physical Society},
  doi = {10.1103/PhysRevLett.133.211803},
  url = {https://link.aps.org/doi/10.1103/PhysRevLett.133.211803}
}

\end{document}